\documentclass{XrU2005}
\usepackage{graphicx}

\title{Emission Measure Distribution and Abundances Measurements During High-T Flares}
\author{R. Nordon}
\affil{Department of Physics, Technion, 32000 Haifa, Israel; nordon@physics.technion.ac.il}
\author{E. Behar}
\affil{Department of Physics, Technion, 32000 Haifa, Israel; behar@physics.technion.ac.il}
\author{M. G\"udel}
\affil{Paul Scherrer Institut, W\"urenlingen \& Villigen, 5232 Villigen PSI, Switzerland; guedel@astro.phys.ethz.ch}

\newcommand{\sg}{$\sigma$ Geminorum}
\newcommand{\cxc}{{\it Chandra}}
\newcommand{\xmm}{{\it XMM-Newton}}

\begin{document}


\maketitle

\begin{abstract}
The RSCVn system \sg\ was observed by \xmm\ on April 2001 during a large flare. We model the emission measure distribution (EMD) during the flare and during a quiescence period. In the flare, a two phase behaviour is found in which the cool plasma ($kT < 2$~keV) is not disturbed while a large hot component, at temperatues of $kT > 3$~keV emerges. Fundamental limitations on EMD modeling of high temperature plasmas are then discussed, in the context of the \sg\ flare.
\end{abstract}

\section{Introduction}

The interplay between steady coronal emission and coronal flares has been a subject of ongoing research for many years. X-ray line resolved spectra available with \cxc\ and \xmm\ allow now for unprecedented plasma diagnostics.

Measurements of relevant changes could offer important diagnostics for the heating and the plasma transport process in stellar coronae. It is unclear how exactly large flares affect the abundances and thermal structure in stellar coronae. Although indications of changes in the metallicity of a flaring corona were reported early on from low-resolution devices \citep[see review by][]{Gudel2004}, more reliable analysis had to wait for the advent of high-sensitivity and medium-to-high resolution spectrometers.

A crucial reconsideration of the situation  came with the advent of high-resolution spectroscopy with {\it XMM-Newton} and {\it Chandra}, but the findings so far still lack systematic trends. While \citet{audard01} found significant enhancements of low-FIP elements in a flare on HR 1099, two flares reported by Osten et al. (2003; for $\sigma^2$ CrB) and G\"udel et al. (2004; for Proxima Centauri) showed an increase of the abundances of several elements, but no selective FIP-dependence was found.

Resolving the thermal structure and finding element abundances are entangled problems that need to be solved simultaneously. The most common approach is the {\it global fit}, where the plasma is treated as composed of several isothermal components. Although this presentation of the thermal structure is not very physical, it usually proves useful for abundance calculations. When the thermal structure itself is of interest, attempts are made to produce a continuus thermal distribution: Emission Measure Distribution (EMD) or sometimes referred to as Differential Emission Measure (DEM). This is achieved by inverting a set of integral equations for emission lines and / or the continuum. The inversion problem however is mathematically ill posed and the solution is unstable against errors in the measurements, {\it ie.} small variations of the measured fluxes translate into large variations of the EMD solution (Craig \& Brown 1976). 

RS~CVn binary systems are bright X-ray and EUV sources owing to rapid rotation, generating a magnetic dynamo. As such, they have been studied extensively in both bands \citep[][respectively, and references therein]{Audard2003, Sanz-Forcada2002}.
The RS~CVn $\sigma$ Geminorum (HD62044, HR 2973, HIP 37629) is particularly bright and well observed at all wavelengths. For an RS CVn, it has a rather long period of 19.6045 days \citep{Duemmler}. The primary star is a K1 III type, red giant. Little is known of the secondary as it has not been detected at any wavelength, but restrictions to its mass and the low luminosity suggest that it is most likely a late-type main sequence star of under one solar mass \citep{Duemmler}.

The \sg\ system is a luminous X-ray \citep[$\log L_X \approx 31.0 \pm 0.2$~erg~s$^{-1}$;][corrected for a distance of 37.5~pc]{Yi} and radio \citep[$\log L_R \approx$ 15.40~erg~s$^{-1}$~Hz$^{-1}$, at 6~cm wavelength;][]{Drake1989} source. While most observations found it to be a relatively steady emitter, a very large flare has been detected in December 1998 with EUVE \citep{Sanz-Forcada2002}. Another flare was detected in April 2001 both in the X-ray and in the radio in which a Neupert effect: $\frac{d}{dt}L_{X} \propto L_{Radio}$ \citep{Neupert1968} was found \citep{Gudel2002}. 

In this paper we will use the \sg\ example to discuss some of the disgnostics problems that originate in the high temperatures of the plasma in large flares. For a more complete analysis of the high resolution \sg\ observations discussed below, see Nordon et al. (2005). 

\section{Observations}

The target \sg\ was observed by \xmm\ in April 2001 for a total exposure time of 54~ks. The data were reduced using the Standard Analysis System (SAS) version 6.0.0.
In this analysis we use the Reflection Grating Spectrometers (RGS) in the 1st order of diffraction, which gives reliable data from 6 to 38~\AA. Line fluxes of Fe$^{24+}$ and Fe$^{25+}$ were extracted from the EPIC-pn data with the use of the XSPEC software package. Background is subtracted using off-source CCD regions. 
\cxc\ observed the target on December 1999 for a duration of 100~ks with the Low Energy Transmission Grating (LETG) + Advanced CCD Imaging Spectrometer (ACIS) configuration in Continuous Clocking (CC) mode. The data were reduced using the CIAO package version 3.0.2.

The long duration flare, with an observed peak rise of 20\%, is seen in the RGS light curve presented in Figure~1. The rise in flux is roughly uniform in the entire RGS wavelength band. The flare light curve is contrasted with the flat light curve obtained from the LETG observation shown below in the same figure.

\begin{figure}
\label{fg:lightcurves}
\centering
\includegraphics[width=1\linewidth]{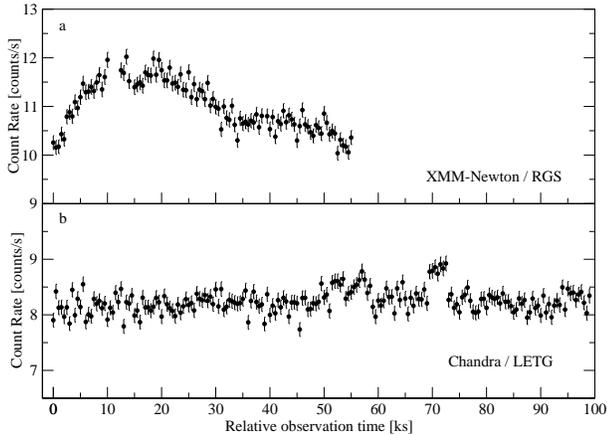}
\caption{Light Curves for \sg\ in time bins of 500~s. (a) April 2001 RGS (1 and 2 combined) observation (1st and 2nd orders). (b) December 1999 LETG/ACIS observation (all orders).}
\end{figure}

\section{Emission Measure Distribution}

\subsection{Emission measure distribution modeling}
The continuum is almost featureless for a distrubution of plasma temperatures and in addition, the continuum from the LETGS instrument is unreliable due to the CC mode (Nordon et al. 2005). Therefore line fluxes alone are used to derive the emission measure distribution (EMD).

The observed line flux $F^{q}_{ji}$ of ion $q$ due to the atomic transition $j \rightarrow i$ can be expressed by means of the element abundance with respect to Hydrogen $A_z$, the distance to the object $d$, the line power $P^q_{ji}$ and the ion fractional abundance $f_q$ as:

\begin{equation}
  F^q_{ji} = \frac{A_z}{4 \pi d^2} \int_{0}^{\infty}{P^q_{ji}(T) f_q(T) EMD(T) \mathrm{d}T }
  \label{eq:lineflux}
\end{equation}

We use the primary line (selected bright, weakly blended, usually resonant line) from every Fe ion in the observed spectra to get a set of integral equations (eq.~\ref{eq:lineflux}), whose solution yields the EMD scaled by the unknown Fe abundance. For other elements, we use ratios of the He-like to H-like line fluxes instead of absolute fluxes, thus the element abundance $A_z$ cancels out. This adds another set of equations that constrain the shape of the EMD, and do not depend on the abundances:

\begin{equation}
  R_z = \frac{ F^{He-like}_{ji} }{ F^{H-like}_{lk} }
  \label{eq:fluxratio}
\end{equation}

The X-ray spectra include as many as ten Fe ions but no more than two ions from other elements. In total, we get 14 equations for the flare and quiescence observations, but different equations, depending on which lines are visible. We fit the line fluxes and flux ratios using the {\it least squares best fit} method to solve for the EMD, where the EMD is expressed by a parameteric non-negative function of $T$. 
This method yields the estimated shape of the EMD, independent of any assumptions for the abundances, and is scaled by the Fe abundance. The integration in eq.~\ref{eq:lineflux} and eq.~\ref{eq:fluxratio} is cut-off at 8~keV beyond which the EMD is completely degenerate. This means that some of the EM in the last bin could be attributed to even higher temperatures.

The atomic data for the line powers are calculated using the HULLAC code \citep{HULLAC}. In order to measure the line fluxes and solve for possible blending, we preform an ion-by-ion fitting to the spectra. The line powers for each ion are calculated at its maximum emissivity temperature and then passed through the instrument response. The observed spectra are fitted by a set of complete individual-ion spectra simultaneously, resulting in an excellent fit that accounts for all the observed lines and blends. This process is similar to the one used in \citet{Behar2001} and \citet{Brinkman2001}.
The line fluxes used in the EMD fitting are listed in \citet{Nordon2005}.
The ionic abundances ($f_q$) for: Fe, Ar, S, Si, Mg are taken from \citet{Gu2003}, whereas \citet{Mazzotta1998} is used for the other elements. 

Our goal is to compare the EMD of the flare and quiescence states. It is important to note that the solution for the EMD is not unique as is the case with integral equations of this sort \citep{Craig1976}. On scales much smaller than the width of the ions emissivity curves, or in temperature regions where there are no emissivity peaks of any ion, there is no way of constraining the EMD. Therefore, in order to be able to compare the EMD solutions, the confidence intervals of the solution are as important as the actual values.
We choose to fit a {\it staircase} shaped function to allow for local confidence intervals estimates. Since the errors on the EM can be non-linear in the parameters (due to the flux ratio equations), localized asymptotic confidence intervals are inappropriate. We use the $\chi^2$ probability distribution to get the confidence intervals, by searching the model parameters space for a target $\chi^2$ surface. The deviation of the target $\chi^2$ from the best-fit value gives the confidence level. This also includes uncertainties due to non-zero covariances between parameters.

The fitted EMD of \sg\ during flare and quiescence are plotted in Figure~2. with 90\%\ confidence intervals. The value of the EMD in each bin represents the average EMD over the bin. Selecting the number of bins and their widths is not trivial. Since, as discussed above, we are interested in meaningful confidence intervals, we cannot use narrow bins, as this will result in excessive error bars. The line emissivity curves have considerable widths and some extend to temperatures much higher than their peak emissivity, resulting in strong negative correlations between the EM in neigbouring bins. Therefore, on small enough scales we have no information on how the EM is distributed and we can only constrain the total EM (or average EMD) over a bin's temperature range. 
The errors on the measured fluxes increase the uncertainty on the EM even further. Ultimately, if meaningful confidence intervals are to be obtained, the number of bins has to be kept small and their width optimized according to the constraints in each region.

\begin{figure}
\label{fg:EMD_plot}
\includegraphics[width=1\linewidth]{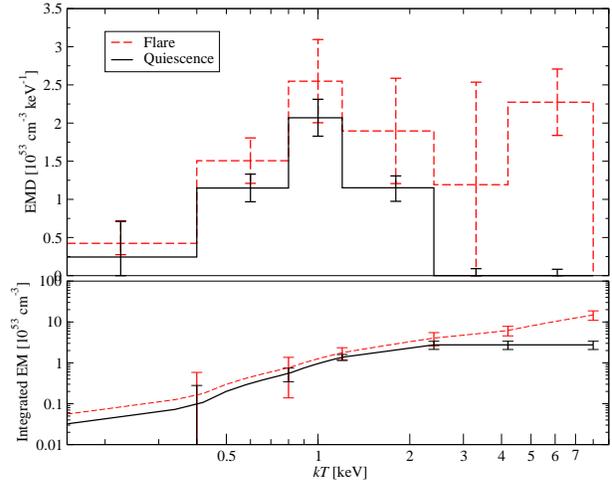}
\caption{{\it Top:} EMD of the two observations. Error bars indicate 90\%\ confidence intervals. {\it bottom:} The integrated EM up to $kT$, with 90\%\ confidence intervals. EMD is scaled according to Solar Fe abundance, taken to be: Fe/H = 4e-5.}
\end{figure}

\subsection{Integrated Emission Measure}

The important physical quantity is the {\it integral} of the EMD over a range of temperatures. 
The integrated EM from zero to $kT$ is plotted in the bottom panel of Figure~2. with 90\%\ confidence bars. The uncertainties caused by the strong correlation between the EMD bins disappear with integration, resulting in much smaller error bars. Several works comparing different EMD reconstruction methods showed that while the representation of the EMD is heavily model dependent, the total integrated EM stays remarkably the same (for example, Fludra \& Sylwester, 1986). An important exception to this is when significant amounts of EM exist at temperatures where all line emissivities are low, such as beyond the emissivity peaks of the highest Fe degrees. 

Simple heating of the plasma does not change the total EM, therefore variations in the total EM indicate added material, changes in density or both. 
The total integrated EM up to 3~keV is slightly higher during the flare, although still consistent with the quiescence EM, within the 90\%\ confidence intervals.
The total EM up to the 8~keV cut-off during the flare is 5.4$\pm$1.9 times that of the quiescence EM, which is very large in a bright system such as \sg, so it is unlikely that such a huge amount of plasma is added to the corona. The more likely interpretation is that the hot EM originates from plasma heated lower in the chromosphere, where the higher density would result in a large EM, even for a small amount of evaporated plasma. 
The increase in density was not detected here, but this could be due to the high charge states (H-like and bare) typical of the high-T flare, for which no density diagnostics are available.

\subsection{Abundances}

Once we have solved for the EMD, abundance calculations are made easy.
In order to extract the X/Fe abundance ratios, we simply calculate the non-Fe line fluxes (eq.~\ref{eq:lineflux}) from the Fe-scaled EMD. The ratio between the measured and calculated flux gives the abundance value. We use abundance ratios relative to Fe and not to H since this is based on line emission ratios, while H does not emit lines at the relevant temperatures and abundances relative to H have to rely on ambiguous continuum modeling.

No statistically significant abundance variations between flare and quiescence were detected. This does not necessarily mean that the abundances are the same since the high temperatures in the flare can make the detection of abundance variations difficult, as we explain in section 5.

\section{EMD Degeneracy and Confidence Level}
\subsection{Definitions}
The inversion of a set of integral equations as presented above is not unique. Even for very small errors on the measured fluxes, we get an infinite number of possible solutions (in the {\it least squares sense}). 
Therefore, estimating the EMD uncertainties is crucial for the physical interpretation of the EMD and for comparisons between different activity epochs.
It is important to note that the EMD uncertainty in a given $T$ range is caused by a degeneracy of the inversion problem itself and not merely by the errors on measured fluxes, as we will explain below. This is not a problem caused by a model. The goal is to find what variations on the {\it source} EMD will result in indistinguishable spectra and to determine the maximal temperature resolution which can be achieved, regardless of our specific EMD representation or method of solution.

We would like to define a measure for the degeneracy of the EMD solution at a given temperature T. For simplicity of notation let us define $\epsilon_i(T)$ as the emissivity of an observed line $i$:

\begin{equation}
  \epsilon_i (T) = \epsilon^q_{kl}(T) \equiv  P^q_{kl}(T) f_q(T)
  \label{eq:eps_definition}
\end{equation}

\noindent Using this definition, the observed line flux (eq.~\ref{eq:lineflux}) can be written as:

\begin{equation}
  F^{i} = \frac{A_Z}{4\pi d^2} \int_{0}^{\infty}{ \epsilon_i (T) EMD(T) \, \mathrm{d}T }
  \label{eq:Fi}
\end{equation}

\noindent $\epsilon_{i}(T)$ defines the contribution of the plasma at temperature interval $[T, T+\mathrm{d}T]$ to each observed line flux $F^{i}$ scaled by $EMD(T)\mathrm{d}T$. The EMD contributions at a given temperature $T$ to all of the observed line fluxes $F^{i}$ ($i$=1..N, number of observed lines) forms a vector $\vec{\epsilon}(T)$. 

Let us now quantify the similarity between the emissivity vectors at different $T$. In other words, how much plasma at $T{_1}$ can be replaced with plasma at $T_2$ and still produce the same observed line fluxes within the observational errors.
Varying the EMD at $T_1$ by $\delta EMD(T_1) = A_1 \delta(T-T_1)$, results in a deviation in the line flux vector $\vec{F}$. This deviation can be compensated partially by another EMD variation at temperature $T_2$, $\delta EMD(T_2) = A_2 \delta(T-T_2)$. If vector $\vec{\epsilon}(T_2)$ were parallel to $\vec{\epsilon}(T_1)$, there would be total compensation. Therefore we would not distinguish between plasma at $T_1$ and plasma at $T_2$. 
In the general case, the deviation in the observed line fluxes is:

\begin{equation}
 \frac{4\pi d^2}{A_Z} \Delta \vec{F} = A_2 \vec{\epsilon}(T_2) - A_1 \vec{\epsilon}(T_1) 
  \label{eq:delta_F}
\end{equation}

\noindent The condition for $|\Delta \vec{F}|$ to be minimal, is that $A_2$ satisfies:

\begin{equation}
  A_2 |\vec{\epsilon}(T_2)| = A_1 |\vec{\epsilon}(T_1)| cos \theta _{1,2}
  \label{eq:min_variation}
\end{equation}

\noindent where $\theta _{1,2}$ is the angle between the emissivity vectors of temperatures $T_1$ and $T_2$:

\begin{equation}
  \cos \theta _{1,2} = \frac{ \vec{\epsilon}(T_1) \cdot \vec{\epsilon}(T_2) }{ |\vec{\epsilon}(T_1)| |\vec{\epsilon}(T_2)| }
  \label{eq:cos_theta}
\end{equation}

\noindent The {\it correlation angle} $\theta_{1,2}$ is a measure of the spectral similarity of plasmas at different $T$. In practice one needs to weigh flux deviations by the errors on the measured line fluxes $\sigma_i$. We can define a modified emissivity vector $\vec{\xi}(T)$ whose components are $\xi_i(T) = \epsilon_i(T) / \sigma_{i}$. An appropriate {\it correlation angle} $\tilde{\theta}_{1,2}$ can then be defined:

\begin{equation}
  \cos \tilde{\theta}_{1,2} = \frac{ \vec{\xi}(T_1) \cdot \vec{\xi}(T_2) }{ |\vec{\xi}(T_1)| |\vec{\xi}(T_2)| }
  \label{eq:corr_angle}
\end{equation}

\noindent Consequently, we define a simple expression for a $\chi^2$-like significance that represents the ability to discriminate between the EMDs that produce the two sets of line fluxes:

\begin{equation}
  \Delta \chi^2_{T_1, T_2} = |A_1 \vec{\xi}(T_1) - A_2 \vec{\xi}(T_2)|^2
  \label{eq:delta_chi_sqr_2_def}
\end{equation}

\noindent Now, the minimization of $\Delta \chi^2$ requires that:

\begin{equation}
  \Delta \chi^2_{T_1, T_2} = A_1^2 |\vec{\xi}(T_1)|^2 \sin^2 \tilde{\theta}_{1,2}
  \label{eq:delta_chi_sqr_2}
\end{equation}

We can use $|\vec{\xi}(T_1)| \sin \tilde{\theta}_{1,2}$ to define and map a {\it degeneracy length} which will set a minimum temperature width to features we can detect in the EMD, with respect to their EM ($A_1$). The limit is set by:
\begin{equation}
  A_1(T_1, T_2) = \frac{ \sqrt{\Delta \chi^2} }{ |\vec{\xi}(T_1)| \sin \tilde{\theta}_{1,2} }
  \label{eq:A1}
\end{equation}

Any EMD variations of am amount $A_1$ or less of total EM inside this temperature range will produce the exact same spectra up to the accuracy of measurement, quantified by $\Delta \chi^2$. This approach can be expanded to include also the continuum by measuring it at different wavelengths and treating the flux at those wavelengths as lines. However, we do not use the continuum here since it was not used in the \sg\ analysis. 

\begin{figure}
  \begin{center}
    \includegraphics[width=\linewidth]{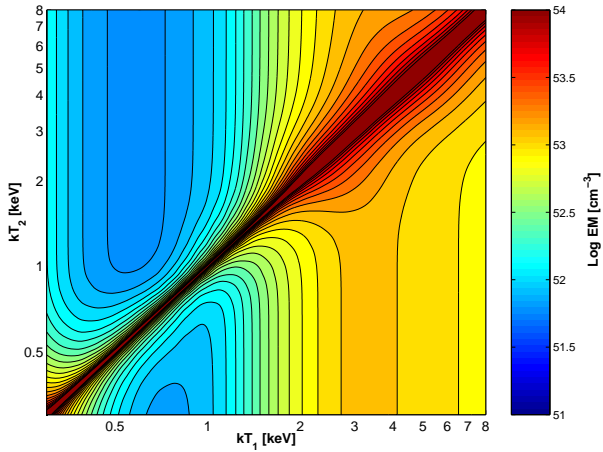}
    \caption{Constraints map for the EM using only lines of Fe$^{16+}$ to Fe$^{25+}$. Colour scale is $\log A_1$. Selecting a temperature $T_1$ the contour lines mark the temperature uncertainty $T_2$ for the corresponding EM. Around 1~keV we get good constraints due to Fe$^{16+}$ to Fe$^{23+}$ having narrow emissivity curves that peak between 500 and 1600 eV. The problematic region around 3~keV is marked by contour lines stretching outside of the map.}
    \label{fg:corr_mapping}
  \end{center}
\end{figure}

\subsection{Application to \sg}
Setting $\Delta \chi^2 = 1$ and using only Fe lines, where flux errors are taken from the flare observation in \citet{Nordon2005}, the mapping of $A_1(T_1, T_2)$ is plotted in Figure~\ref{fg:corr_mapping}. We note that $A_1$ tends to infinity at $T_1=T_2$.

\begin{figure}
  \begin{center}
    \includegraphics[width=\linewidth]{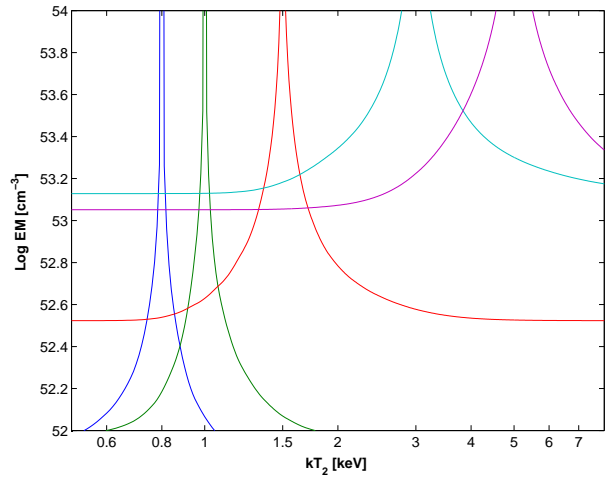}
    \caption{Cross sections of Figure~\ref{fg:corr_mapping} at various constant $kT_1$ values. Note that $\log$ EM tends to infinity at $kT_1=kT_2$. }
    \label{fg:crossection}
  \end{center}
\end{figure}

In order to make reading and understanding of Figure~\ref{fg:corr_mapping} easier, cross sections of the map at fixed $T_1$ are plotted in Figure~\ref{fg:crossection}.
The width of a curve at a given EM is the fundamental uncertainty in temperature for a component of that size. For example, the crossection at $kT=1$~keV shows that an EM amount of $\log EM < 52.5$~cm$^{-3}$ from this temperature may be redistributed within an interval of $\pm$0.1~keV and still produce the same spectra in the sense of $\Delta \chi^2 \leq 1$. Therefore using bins of less than 0.2~keV width around this temperature is meaningless if the average EMD $<$1.5 10$^{53}$~cm$^{-3}$ keV$^{-1}$. In the case of \sg, the average EMD at $kT=1$~keV was only slightly higher $\sim$2.2 10$^{53}$~cm$^{-3}$keV$^{-1}$, which will result in EMD uncertainty making it almost consistent with zero. Eventually a 0.8-1.2~keV bin was used in Figure~2 in order to have much more EM in the bin than this uncertainty value. Using averaged EMD over a wider bin allows us to put meaningful, well {\it localized} constraints on the EMD at the cost of temperature resolution.

As temperature rises the constraint generally becomes weaker and bins containing more total EM are needed. In Figure~\ref{fg:corr_mapping} the problematic region around 3~keV, where there is a gap between the emissivity peaks of Fe$^{23+}$ and Fe$^{24+}$, is clearly visible by the bend in the contour lines. This is the reason the 3~keV bin in Figure~2 is poorly constrained. 

One has to remember that the mapping of $A_1(T_1,T_2)$ is observation-specific, as it depends on the line flux errors. Still, we can conclude that generally: The fundamental degeneracy is caused by the slow variation with $T$ of the emissivity vector $\vec{\epsilon}(T)$. The scale of the degeneracy problem is set by the absolute value of the {\it modified} emissivity vector $|\vec{\xi}(T_1)|$ which can be understood as a signal-to-noise ratio per EM unit. 

\section{Difficulties of Hight-T Flares}

\begin{figure}
  \begin{center}
    \includegraphics[width=3.25in]{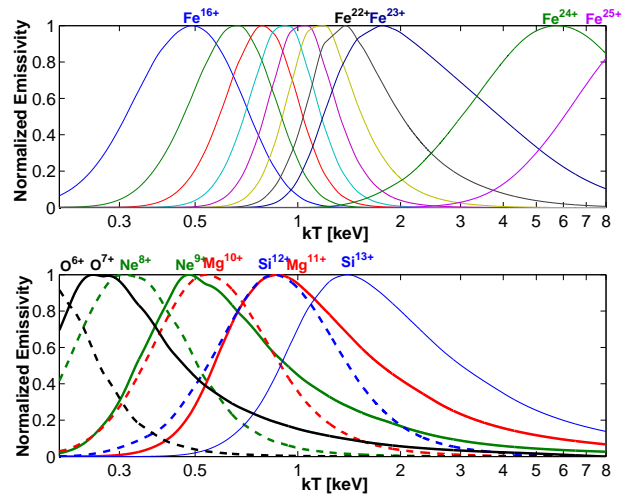}
    \caption{Emissivity of RGS lines used in the flare EMD reconstruction. {\it Top:} Lines of the various Fe ions. {\it Bottom:} Lines of other elements, dotted lines are He-like ions, solid lines are H-like ions. Note the lack of emissivity between 2 and 4~keV, the narrow emissivity curves of lower Fe ions compared to the other elements and the emissivity curve of Mg$^{11+}$ stretching out to high T. Si$^{13+}$ is on the edge of the RGS instrument and suffers from very bad signal to noise ratio, so its contribution to constraints on the EMD is minimal. }
    \label{fg:emissivities}
  \end{center}
\end{figure}

When a large amount of EM is located at high temepratures, as is often the case in large flares on RSCVn systems,  several difficulties in measuring plasma parameters arise. When temperatures are high enough, a significant portion of the elements are in their H-like ionic form or worse - bare, and line diagnostics lose power. Our ability to measure densities relies mostly on He-like triplets. However, in a two-phase scenario like in \sg, what we measure in practice is the density of the surrounding cooler quiescent plasma and not the density in the flaring region where there are no He-like ions. In Figure~\ref{fg:emissivities} we plot the normalized emissivities of lines available in the RGS (plus the He-like \&\ H-like Fe from EPIC) during the flare. We see that the emissivity of He-like ions (dashed lines in lower panel) is extremely low above 2~keV.
The same problem applies to abundances. For example, a high abundance of oxygen in a $kT=6$~keV region is invisible to us since at 6~keV oxygen is in the bare state (but we will still detect the undisturbed surrounding cooler plasma).

We can also see from Figure~\ref{fg:emissivities} that the kT range of 0.5-2~keV is well covered by eight Fe ionization degrees and most of the other elements. Emission from higher kT is covered only by the highest degrees of Fe - mostly Fe$^{24+}$ and very weak tail-end emission of H-like ions of other elements, which leads to higher uncertainties in the EMD, as discussed in the previous section. It is also interesting to note that the emissivity of Mg$^{11+}$, although peaking at $\sim$850~eV stretchs out to temperatures far beyond the peak and into the range of the flare temperatures. In the RGS it will be the only non-Fe ion to be effected from EM above 2~keV as Si$^{13+}$ usually suffers from high noise and is not useful.

In this \sg\ example we determined the flaring plasma to be mostly above 3~keV. Therefore, all density measurements and element abundances of C,N,O,Ne,Mg and Si (available in the RGS) during the flare are in essence, measurements of the surrounding quiescent plasma. In this observation we do not have post-flare data since the observation ended before the flare fully decayed, but in such a case it will be interesting to measure the abundaces immediately after the flare, when most of the flaring plasma has cooled to quiescent temperatures. Still, if the large EM of the flare originated in the higher densities of the chromosphere, the evaporating plasma will rise and cool in the less-dense corona. Hence, the EM will drop significantly and we might not be able to detect variations on the significant background emission of the surrounding plasma unrelated to the flare.

In conclusion, results of previous works, where some detect density and abundance variations in large flares and some do not, may depend heavily on the temperatures in the flare and the amount of surrounding quiescent plasma, and not only on what actually happened in the flare. Variations in the thermal structure of the plasma are elusive due to the difficulty of determining the uncertainties of possible EMD solutions on small scales. 

\vspace{-0.5cm}
\section*{Acknowledgments}
The research at the Technion was supported by ISF grant 28/03 and by a grant from the Asher Space Research Institute. PSI astronomy has been supported by the Swiss National Science Foundation (grant 20-66875.01). \xmm\ is an ESA science mission with instruments and contributions directly funded by ESA member states and the USA (NASA).


\end{document}